\documentstyle[eqsecnum,floats,preprint,epsf,aps]{revtex}
\tighten

\begin{document}

\renewcommand{\arraystretch}{1.5}
\newcommand{\be}{\begin{equation}}
\newcommand{\ee}{\end{equation}}
\newcommand{\bea}{\begin{eqnarray}}
\newcommand{\eea}{\end{eqnarray}}
\newcommand{\la}{\leftarrow}
\newcommand{\ra}{\rightarrow}
\newcommand{\lr}{leftrightarrow}
\newcommand{\La}{\Leftarrow}
\newcommand{\Ra}{\Rightarrow}
\newcommand{\Lr}{\Leftrightarrow}
\def\Tr{\mathop{\rm Tr}\nolimits}
\def\mapright#1{\smash{\mathop{\longrightarrow}\limits^{#1}}}
\def\mapdown#1{\big\downarrow \rlap{$\vcenter
  {\hbox{$\scriptstyle#1$}}$}}
\def\Z{{\bf Z}}
\def\R{{\bf R}}
\def\M{{\cal M}}
\def\L{{\cal L}}
\def\c{c_\chi}
\def\s{s_\chi}
\def\xh{\hat x}
\def\yh{\hat y}
\def\zh{\hat z}
\def\vt{\widetilde{v}}
\def\wt{\widetilde{w}}
\def\n#1{{\hat{n}_{#1}}}
\def\c#1{\cos{#1}}
\def\s#1{\sin{#1}}
\def\cs#1{\cos^2{#1}}
\def\ss#1{\sin^2{#1}}
\newcommand{\PSbox}[3]{\mbox{\rule{0in}{#3}\includegraphics{#1}\hspace{#2}}}

\title{Charged False Vacuum Bubbles and
the AdS/CFT Correspondence}

\author{Gian Luigi Alberghi$^{1}$, David A. Lowe$^{2}$ and Mark Trodden$^{3}$}

\address{~\\$^1$Dipartimento di Fisica, Universita' di Bologna and \\
I.N.F.N, Sezione di Bologna \\
Bologna, Italy. \\
{\tt alberghi@bo.infn.it}}

\address{~\\$^2$Department of Physics\\
Brown University \\
Providence, RI 02912, USA. \\
{\tt lowe@het.brown.edu}}

\address{~\\$^3$Department of Physics \\
Case Western Reserve University \\
10900 Euclid Avenue \\
Cleveland, OH 44106-7079, USA. \\
{\tt trodden@erebus.cwru.edu}}

\maketitle

\begin{abstract}
We initiate a study of cosmology within the framework of Maldacena's
AdS/CFT correspondence. We present
a comprehensive analysis of the classical motion of a charged domain wall
that separates an external Reissner-Nordstr\o m region of spacetime
(with small or vanishing cosmological constant) from an 
internal de-Sitter region. The possible associated spacetime
diagrams are drawn, although in the classical case, an unambiguous prediction
of what occurs at late times in the interior region is not possible, since
singularities and Cauchy horizons form.  We argue that, when 
the asymptotic region is anti-de Sitter, the AdS/CFT correspondence
gives a prescription for resolving the curvature singularities and
evolving solutions across the expected Cauchy horizon. 
Some of our solutions contain inflating interiors, and we provide
evidence these can be patched onto solutions with smooth initial data, 
circumventing an obstacle found by Farhi and Guth to creating an
inflating universe in the laboratory.

\end{abstract}

\setcounter{page}{0}
\thispagestyle{empty}

\vfill

\noindent BROWN-HET-1186\hfill 

\noindent CWRU-P22-99


\eject

\vfill

\eject

\baselineskip 20pt plus 2pt minus 2pt

\section{Introduction}
\label{sec:introduction}

The conjectured correspondence between large $N$ superconformal field
theories and string theory in anti-de Sitter backgrounds
\cite{maldacena}, 
provides us with a possible nonperturbative definition of string
theory. The formulation is background dependent, since
one is always left with the restriction that the geometry is
asymptotically anti-de Sitter space. Nevertheless it is possible to
use this framework to address some of the longstanding questions in
cosmology by considering processes where a black hole forms with an
interior resembling a realistic cosmological model.

In this paper we take a first step in this direction. We perform a 
comprehensive analysis of the motion of a charged domain wall that separates
an external Reissner-Nordstr\o m  
region of spacetime (with small or vanishing cosmological constant) 
from an internal de-Sitter
region.  
Our analysis rests on the discontinuity equations of Israel \cite{israel}
and Kuchar \cite{kuchar}, and is very much in the spirit of that carried out by
Blau, Guendelman and Guth, for the case in which the external metric is
Schwarzschild \cite{BGG} (see also \cite{berezin}), 
and by Boulware \cite{boulware}, for the case in 
which the internal metric is Minkowski.

In the present study, with a charged shell, there
exists a much richer spectrum of possibilities for the dynamics of the 
collapsing shell. We avoid an obstacle
encountered in \cite{FG 87} that initial conditions are
necessarily singular.
We present all possible allowed dynamics for the shell in four
dimensions. In the situation considered here,
it is impossible to predict, without imposing additional boundary
conditions,
the ultimate
outcome of the interior region of such a composite spacetime \cite{FG 87},
since the singularity theorems of 
General Relativity predict a singularity to the
past of the core region. 

However, this lack of predictability can be dealt with by
appealing to the AdS/CFT correspondence.
Taking the point of view that this
conjecture is correct, it becomes possible to resolve the
singularities that appear
and unambiguously determine
the dynamics.
Thus, we 
discuss 
the gravitational
systems explored here in the framework of the AdS/CFT correspondence.

This study also addresses a number of issues important for topological
inflation \cite{topinf},
where super-heavy magnetic monopoles are argued to provide the seeds for
inflation. In addition, when 
solitons of various different topological charge are present in one theory 
\cite{DTV 98}, it has been suggested that monopole collisions could give rise 
to inflating core regions \cite{BTV}. The lack of predictability at
the classical level mentioned above is a severe obstacle in these approaches.

The structure of the paper is as follows. In the next section we derive the 
equations of motion of our charged shell.  In section \ref{sec:trajectories} we
then discuss the possible solutions to these equations, and present
the trajectories on
Penrose diagrams. 
Finally, in section \ref{sec:adscft} we discuss the bubble solutions
using the AdS/CFT correspondence and comment on their
interpretation at the quantum level.

\section{Equations of motion}
\label{sec:eom}

Consider a spherical domain wall in 
$n+1$ spacetime dimensions which separates an interior region of spacetime
from an exterior region, described respectively by the static metrics, 
in spherical polar coordinates 
$(r,\Omega)$,
\be
   ds^2 = - f_{in(out)}\;(r)\;dt^2 + f^{-1} _{in(out)}\;(r) \;dr^2 + r^2 
       \;d \Omega_{n-1} ^2~.
\ee
In particular we will examine the case where the internal region is
in a false vacuum state described by de Sitter metric 

\be
\label{fin}
   f_{in} = 1 - \chi^2 r^2~,
\ee
and the external is described by Reissner-Nordstr\o m-anti de Sitter (R-N) 
metric

\be
\label{fout}
   f_{out} =   1 - {2m \over r^{n-2}} + {Q^2 \over r^{2n-4}}  + \Lambda r^2 \ ,
\ee
where $ m $ is the Schwarzschild parameter, $  Q $ the charge, and
$ \Lambda $ the absolute value of the cosmological constant. 
Further, we will only consider the case $ Q \le m $, to preserve the horizon
structure of the R-N black hole,
and  $ \Lambda $ much less than any other parameter, to
ensure a relevant contribution only in the
asymptotic region $ r \ra \infty $.

We will parameterize the radius $r$ of the shell by the proper time $ \tau $
measured by an observer comoving with the shell.  The equations of motion 
are referred to as {\it junction equations} \cite{israel}, and in our 
parameterization reduce to

\be
\label{motion}
    s_{in} \sqrt{ f_{in}+ \dot r ^2} - s_{out}  \sqrt{ f_{out}+ \dot r ^2}
    = k r \ ,
\ee
where $ k$ is proportional to the tension of the wall,
$ s _{in \; (out)} = \pm 1 $, and a dot represents a derivative with respect
to proper time. Squaring (\ref{motion}) twice, we obtain

\be
\label{eqpot}
   \dot r ^2 + V(r) = -1 \ ,
\ee
which is the equation for the one dimensional motion of a particle in a
potential

\be
 V(r)=
 - \chi^2 r^2 - {1 \over 4 k^2 r^2} 
   \left[ (k^2-\chi^2 -\Lambda )r^2 + {2m \over r^{n-2} } - {Q^2 \over r^{2n-4} } 
\right]^2~.
\ee

It can be shown that in regions where $r$ is a spacelike coordinate
$ s_{in(out)} $ are positive if the outward normal to the wall
is pointing towards increasing radii (see \cite{boulware}), and
negative if the normal points towards decreasing radii.
If we use Kruskal-Szekeres style coordinates $(u,v)$ we find that $
s_{in(out)} $ determine also whether the angle $\arctan (v/u)$ increases or 
decreases as we move along the trajectory (see \cite{BGG}).
For the external metric the 
Kruskal-Szekeres coordinates have a rather complicated relationship with
$r$ and $t$. For definiteness we give the explicit relations for
$n=3$, $\Lambda=0$ in region $I $
($u>0$, $u>|v|$) when the formulas are still reasonably simple
\be
\label{kkrn}
u = e^{\gamma r_* } \cosh( \gamma t)~,
\qquad
v = e^{\gamma r_*} \sinh( \gamma t)~,
\ee
where we have defined
\be
\label{tort}
r_* = r + {r_+^2 \over {r_+-r_-}} \log(r-r_+) -{ r_-^2 \over
  {r_+-r_-}} \log(r-r_-)~, \qquad
\gamma = {{r_+-r_-} \over 2r_-^2}~,
\ee
and
where $r_+$ and $r_-$ denote the positions of the 
outer and inner event horizons of the R-N geometry
respectively.
$s_{out}=+1$ implies $\arctan (v/u)$ increases along the trajectory,
while $s_{out}=-1$ implies the angle decreases. For the internal de Sitter
metric, the Kruskal-Szekeres style coordinates in region $I$ ($u>0$, $u>|v|$) are
\be
\label{kkds}
u= \left( { 1-\chi r\over 1+\chi r} \right)^{1/2} \cosh(\chi t)~, \quad
v= \left( { 1-\chi r\over 1+\chi r} \right)^{1/2} \sinh(\chi t)~.
\ee
In this case, $s_{in}=+1$ implies $\arctan (v/u)$ 
decreases along the trajectory,
while $s_{in}=-1$ implies the angle increases.
We now use these relationships to determine the
regions of the Penrose diagrams of the 
internal and external spaces in which the wall evolves.

To begin, (\ref{motion}) yields

\be
\label{sin}
    s_{in} = +1 \,  \Leftrightarrow \, f_{in} - f_{out} + k^2 r^2 > 0~,
\ee
and 

\be 
\label{sout}
   s_{out} =+1 \, \Leftrightarrow \, f_{in} - f_{out} - k^2 r^2 > 0 \ .
\ee
These lead to the relations

\bea
\label{boths}
   s_{out} =+1 \, & \Rightarrow & s_{in} =+1 \nonumber \\
   s_{in} =-1 \, & \Rightarrow & s_{out} =-1 \ .
\eea
For convenience, define the following two functions:

\bea
P(r) & \equiv & (\chi^2+\Lambda+k^2)r^{2n-2}-2mr^{n-2}+Q^2 \nonumber \\
N(r) & \equiv & (\chi^2+\Lambda-k^2)r^{2n-2}-2mr^{n-2}+Q^2 \nonumber \ ,
\eea
with asymptotic behaviors

\be
\label{limsout}
  \lim_{r \ra 0} P(r) = Q^2 ~, \qquad
  \lim_{r \ra \infty} P(r) = + \infty~,
\ee
and

\be
\label{limsin}
    \lim_{r\ra 0} N(r) =Q^2 ~,\qquad
    \lim_{r\ra \infty} N(r)= {\rm sign}(\chi^2+\Lambda-k^2) \, \infty \ .
\ee

First focus on $P(r)$. Substituting (\ref{fin}), (\ref{fout}) in
(\ref{sout}) we obtain 

\be
   s_{out} =+1 \Leftrightarrow P(r) < 0~.
\ee
However, $dP/dr =0$ at only one point, $r=\overline r _{out}$, with

\be 
\overline r _{out} = 
\left[ {m (n-2)\over (n-1)(\chi^2+\Lambda+k^2)} \right] ^{1/n} \ ,
\ee
and hence there are two possibilities for the behavior of $s_{out}$:

\be
 P(  \overline r _{out}) > 0  \Ra  s_{out} =-1  \ ,
\ee
for all $r$, or

\be
   P(\overline r _{out}) < 0  \Ra \left\{ \begin{array}{ll}
     s_{out}=+1 & \ \ \ \mbox{$r_{out}^{(1)} < r < r_{out}^{(2)}$} \\
     s_{out}=-1 & \ \ \ \mbox{$r< r_{out}^{(1)}  \ \ , \ \ r> r_{out}^{(2)}$}
\end{array}\right. \ ,
\ee
where $r_{out}^{(1)}$ and $r_{out}^{(2)}$ are the real solutions of
$P(r)=0$.

Now we turn to $N(r)$. Substituting (\ref{fin}) and (\ref{fout}) in 
(\ref{sin}) we obtain

\be
   s_{in} =+1 \ \Lr \ N(r) <0~.
\ee
Because of (\ref{limsin}), there are more possibilities for $s_{in}$ than for 
$s_{out}$.  If $\chi^2+\Lambda-k^2 <0$, then $dN/dr <0 $ for all $r$. 
In this case

\bea
   s_{in} = +1 \  & \Lr & \ r > r_{in}  \nonumber \\
   s_{in} =- 1 \ & \Lr & \ r < r_{in} \ ,
\eea
where $r_{in}$ is the single solution of $N(r)=0$.
However, if $\chi^2+\Lambda-k^2 >0$ then $dN/dr=0$ at only one point, 
$r=\overline r_{in}$, with

\be
\overline r_{in}=
 \left[ {m (n-2)\over (n-1)(\chi^2+\Lambda-k^2)} \right] ^{1/n} \ ,
\ee
and the corresponding possibilities for the behavior of $s_{in}$ are

\be
    N( \overline r_{in}) > 0 \Ra s_{in}=-1  \ ,
\ee
for all $r$, or

\be
    N( \overline r_{in}) < 0 \Ra \left\{ \begin{array}{ll}
     s_{in}=-1 & \ \ \ \mbox{$r < r_{in}^{(1)}\ \ , \ \ r > r_{in}^{(2)}$} \\
     s_{in}=+1  & \ \ \ \mbox{$r_{in}^{(1)}<r<  r_{in}^{(2)}$}\end{array}\right.
\ee
where $r_{in}^{(1)}$ and $r_{in}^{(2)}$ are the real solutions of
$N(r)=0$.

The possibilities enumerated above for the behavior of the sign parameters
$s_{in}$ and $s_{out}$, are precisely what we need to perform an analysis of
the trajectories of our charged domain wall.

\section{Trajectories}
\label{sec:trajectories}
\begin{figure}
  \centerline{\epsfbox{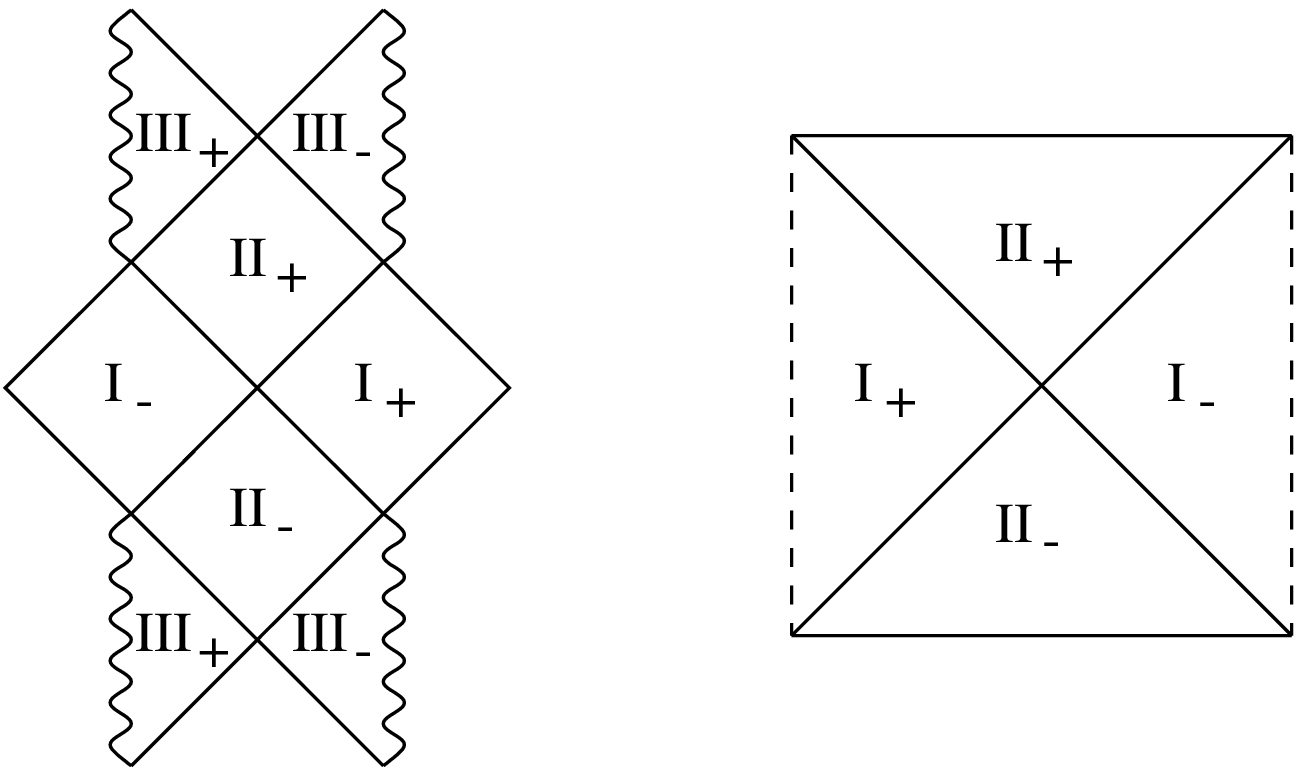}}
  \vspace{0.3in}
  \caption{\sf The separate Penrose diagrams for Reissner-Nordstr\o m (left) 
and
de Sitter (right) spacetimes. }
  \label{originaldiags}
\end{figure}

We will describe the trajectories of the spherically symmetric charged
walls using
Penrose diagrams.  
To keep things simple, we will present diagrams for the familiar
asymptotically flat case. The features of the trajectories we consider 
will be
essentially the same in asymptotically anti-de Sitter space.
The wall traces out a
set of points in a Reissner-Nordstr\o m manifold from the point of view of
an external observer, and a complementary set of points in a de Sitter manifold
from the point of view of an internal observer.  These trajectories
will be represented on pairs of Penrose diagrams describing the
external and internal spaces.
For reference when doing this, and to orient ourselves, we provide 
the separate Reissner-Nordstr\o m and de Sitter Penrose diagrams in 
Fig. (\ref{originaldiags}). Note the Reissner-Nordstr\o m spacetime
contains timelike singularities and a Cauchy horizon, which implies
the classical evolution of the shell is not predictable, without
additional boundary conditions. For the moment, we will simply fix the 
additional boundary conditions by requiring analyticity of the metric
outside the shell, and will comment further on this choice later.

From the discussion of the previous section, it is evident that for a
classically allowed trajectory, the zeros of $N(r)$, $r_{in}$ must satisfy

\be
    r_{in} > r_{H} \equiv 1/ \chi~.
\ee
By considering the physically allowed behavior of the outward pointing 
normal vector of the shell, one finds the zeros of $P(r)$, $r_{out}$ satisfy
\be
  r_{-}<r_{out}< r_{+} \ .
\ee

Now, the time evolution of the shell is completely determined by
equation (\ref{eqpot}) and equations (\ref{sin},\ref{sout}).
The first of these equations determines the global properties of the 
trajectory; in particular whether there exist any inversion points.
In addition, equations (\ref{sin},\ref{sout}) for $ s_{in} $ and 
$ s_{out} $ determine in which regions of the separate Penrose diagrams 
Fig. (\ref{originaldiags}) the trajectory develops.
We will consider only the cases in which the trajectory starts from zero 
or infinite radius.

What are the general features of the trajectories?
First note that equation  (\ref{limsout}) implies that the trajectory
must begin and end in the regions $ I_{-} $ or $ III_{-} $
of the R-N Penrose diagram, while eq. (\ref{limsin}) tells us that
for $ r < r_{H}$ it must be in the right part of the 
de Sitter Penrose diagram. 

Since the equations that determine the turning points of the
trajectories are high order polynomial equations, it is necessary to
solve numerically for the turning points to determine their position
relative to the other features of the trajectory, to check that the
trajectory is physically consistent.
In the remainder of this section we will focus on the case of four
spacetime dimensions ($n=3$), in which the wall
is a two dimensional shell and work out all the possible
trajectories. We have checked that all these trajectories may be
realized also when $n=4$, but will not attempt to check this for
general $n$.

Let us first consider the case in which
\be
\label{unbounded}
   \,\,\,  V(r) < -1~,
\ee
for all values of $r$,
so that there are no inversion points, and in particular let us 
focus on the case in which the shell starts from zero radius and 
expands to infinity. We will call these ``growing'' trajectories.

In this situation there are 5 possible trajectories depending on
the values of $ m$, $Q$, $\chi$, $k$.

\begin{figure}
  \hspace{1.5cm}
  \centerline{\epsfbox{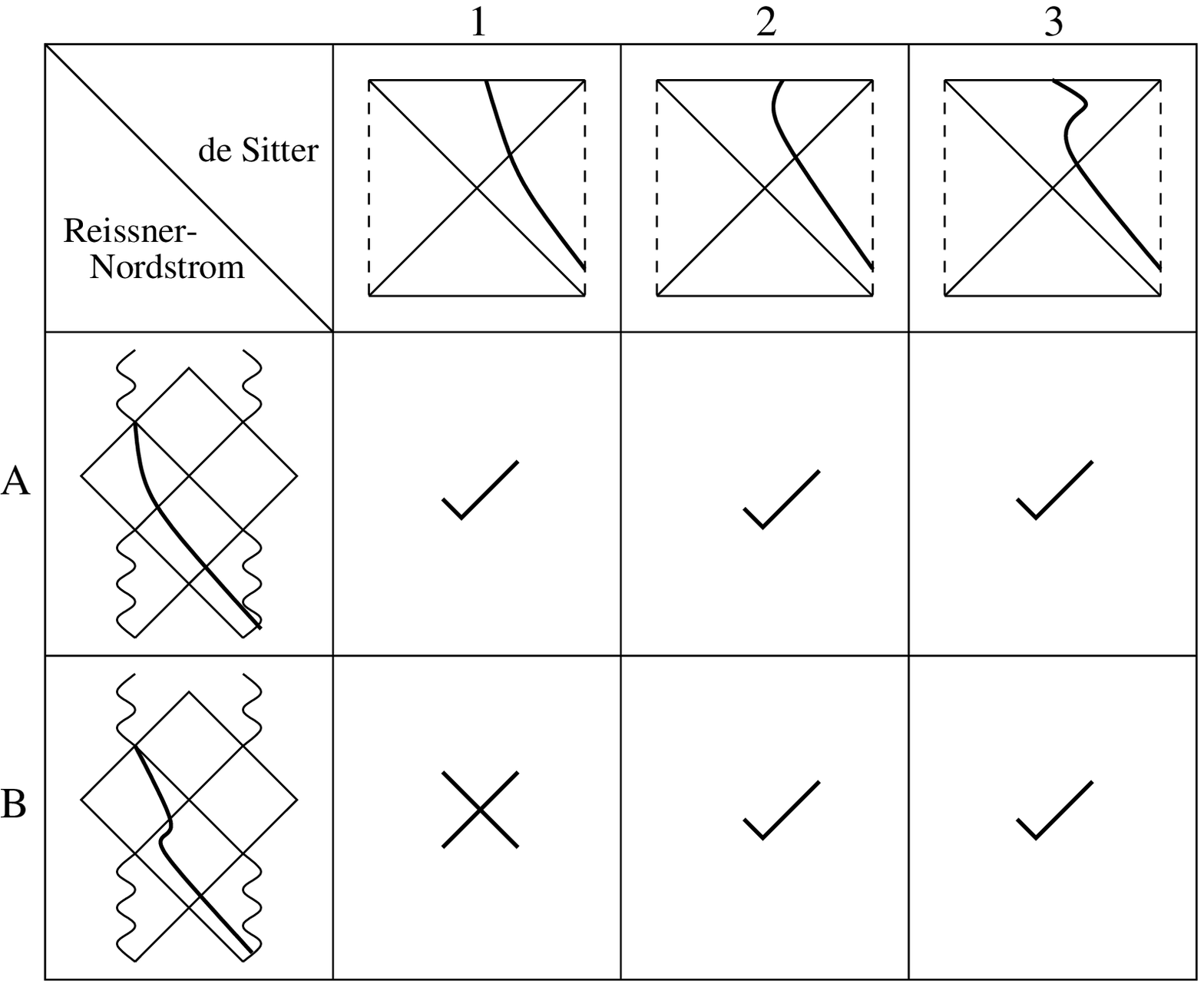}}
  \vspace{0.3in}
  \caption{\sf ``Growing'' trajectories}
  \label{tablegrow}
\end{figure}

\begin{enumerate}
\item {\bf \underline{ $ \chi^2+\Lambda > k^2 \,\, ; \,\,  
     N( \overline r_{in}) > 0 
     \,\,\,\, P(  \overline r _{out}) > 0$}}
\\

In this case $ s_{in}= s_{out}= -1 $ during the whole evolution.

In the de Sitter diagram the shell starts in the right-hand 
patch with an outward normal pointing towards decreasing radii,
and reaches infinite radius keeping always an increasing angle.
In the R-N diagram the trajectory starts in a $ III_{-} $ region
($ s_{out} =-1 $ so that the outward normal points towards decreasing radii) 
then crosses $ r_{-} $ and  evolves with a decreasing 
angle to end in a $ I_{-}$ region at infinity.

This behavior is summarized by the Penrose diagrams in square A1 of
Fig. (\ref{tablegrow}).


\item {\bf \underline{$  \chi^2+\Lambda<k^2
\,\,\, ; \,\,\, 
  P(  \overline r _{out}) > 0$}}
\\

In this case $ s_{in} $ changes sign once, at $ r=r_{in} $, while
$ s_{out} =-1 $ for the whole trajectory.
In the de Sitter diagram the shell starts on the right patch 
with zero radius and evolves with an increasing angle until it reaches 
$ r_{in} $ turning then to the right.
In the R-N diagram the shell has the same behavior as in the preceding case.

This behavior is summarized by the Penrose diagrams in square A2 of
Fig. (\ref{tablegrow}).


\item {\bf \underline{$ \chi^2 +\Lambda > k^2
 \,\,\, , \,\,\,    
  N( \overline r_{in}) < 0 \,\,\, ; \,\,\,  
 P(  \overline r _{out}) > 0$}}
\\

In this case $ s_{in} $ changes twice, at $ r_{in}^{(1)} $ and 
at $ r_{in}^{(2)}$, while $ s_{out} $ never changes.
In the de Sitter diagram the evolution is the same as the previous trajectory
until the first turn at  $ r_{in}^{(1)} $ which is followed by a 
second turn at $ r_{in}^{(2)} $ which lead the shell 
towards the left. The behavior in the R-N diagram is the same as 
in the previous case.

This behavior is summarized by the Penrose diagrams in square A3 of
Fig. (\ref{tablegrow}).


\item {\bf \underline{$ \chi^2+\Lambda > k^2
\,\,\, , \,\,\, N( \overline r_{in}) < 0 
 \,\,\, ; \,\,\,  P(  \overline r _{out}) < 0  $}}
\\

In this case $ s_{in} $ changes at $ r_{in}^{(1)} $  and
$ r_{in}^{(2)} $ and $ s_{out} $ changes at $ r_{out}^{(1)} $ and  
$ r_{out}^{(2)} $.
The behavior in the de Sitter digram is the same as the preceding case.
In the R-N diagram the trajectory starts in the $ III_{-} $ region,
crosses $ r_{-} $, turn to the right at  $ r_{out}^{(1)} $
and to the left as it reaches  $ r_{out}^{(2)} $, to end in 
the $ I_{-} $ region at infinite radius.

This behavior is summarized by the Penrose diagrams in square B3 of
Fig. (\ref{tablegrow}).


\item {\bf \underline{$ \chi^2+\Lambda<k^2
\,\,\, ; \,\,\, 
 P(  \overline r _{out}) <0 $}}
\\

In this case $ s_{in} $ changes one time and $ s_{out} $
twice.
The behavior  is of the same kind as case 
(2) in de Sitter diagram and as the previous case in the R-N spacetime.

This behavior is summarized by the Penrose diagrams in square B2 of
Fig. (\ref{tablegrow}).


\end{enumerate}

Let us now consider the possible trajectories that start at an infinite radius
and collapse to zero, and for which (\ref{unbounded}) still holds.
Once again there are 5 different possible trajectories,
one corresponding to each of those above.

The common features for these trajectories are 
that in the de Sitter diagram the 
trajectories end in the right hand patch with an outward normal
pointing towards decreasing radii while in the R-N diagram the
trajectory starts in a $ I_{-} $ region and ends in a $ III_{-}$ region.

\begin{figure}
  \hspace{1.5cm}
  \centerline{\epsfbox{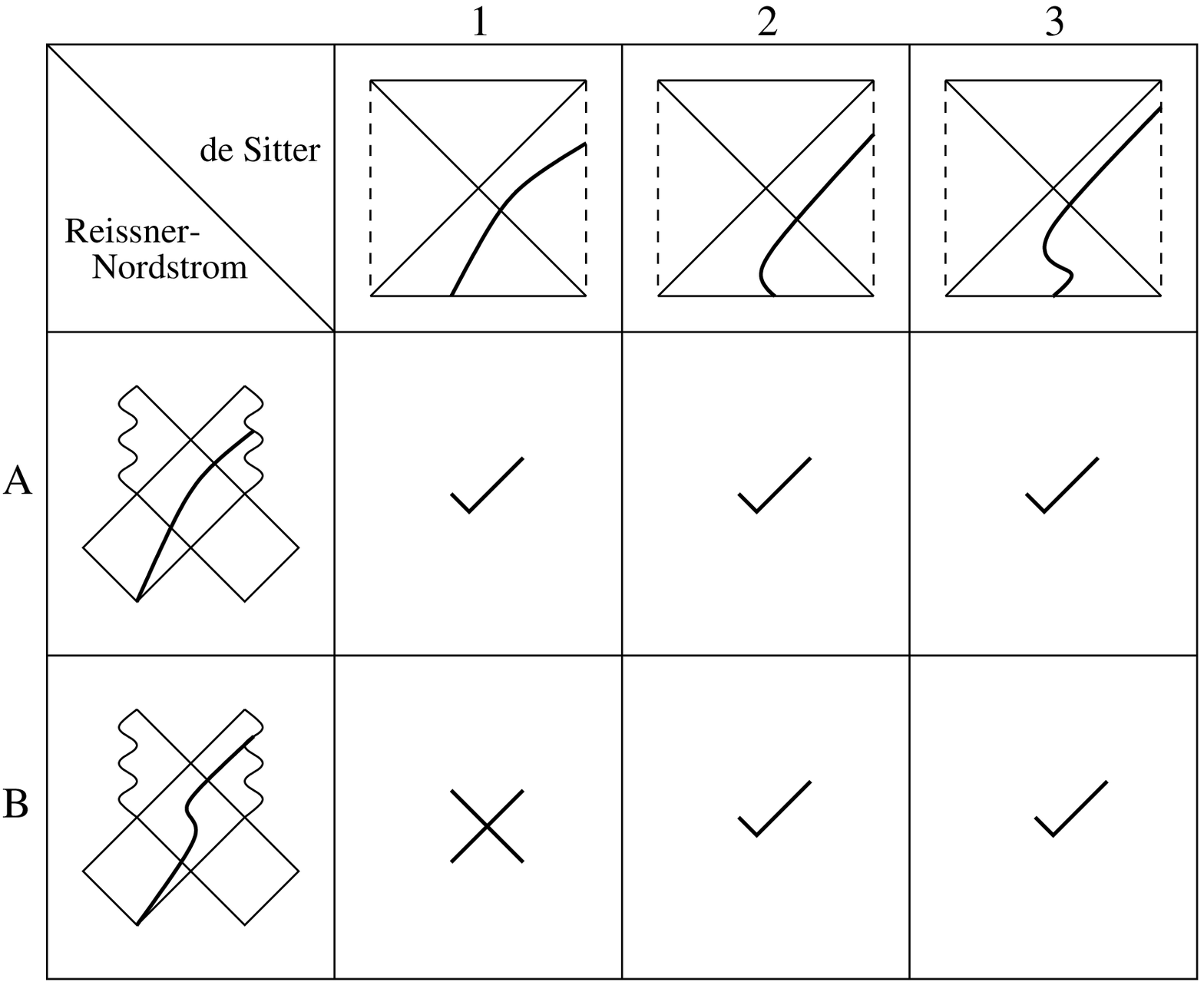}}
  \vspace{0.3in}
  \caption{\sf ``Collapsing'' trajectories}
  \label{tablecoll}
\end{figure}

\begin{enumerate}

\item {\bf \underline{ $ \chi^2+\Lambda > k^2 
\,\, ; \,\,  
     N( \overline r_{in}) > 0 
     \,\,\,\, P(  \overline r _{out}) > 0$}}
\\

In the de Sitter diagram the shell starts from infinite radius with 
an increasing angle ($ s_{in}=-1$) and ends in the right patch with an
outward normal pointing towards decreasing radius.
In the R-N diagram the trajectory starts from a $ I_{-} $ region
and ends in the $ III_{-} $ region without any turning point, since
$ s_{out} = -1 $ throughout the evolution.
This is shown by the Penrose diagrams in square A1 of Fig. (\ref{tablecoll}).



\item {\bf \underline{$  \chi^2+\Lambda<k^2 
\,\,\, ; \,\,\, 
  P(  \overline r _{out}) > 0$}}
\\

In the de Sitter diagram the shell starts at infinity with decreasing angle,
turns at $ r_{in} $ and ends in the right patch.
In the R-N diagram the shell behaves as in the previous case having
$ s_{out} = -1 $ for the whole evolution.
This is shown by the Penrose diagrams in square A2 of Fig. (\ref{tablecoll}).


\item {\bf \underline{$ \chi^2 +\Lambda > k^2 
\,\,\, , \,\,\,    
  N( \overline r_{in}) < 0 \,\,\, ; \,\,\,  
 P(  \overline r _{out}) > 0$}}
\\

In the de Sitter diagram the shell starts at infinity with decreasing
angle, turns at $ r_{in}^{(1)} $ and $ r_{in}^{(2)} $ to end in
the right patch. In the R-N diagram the behavior is the same as in 
the previous case.
This is shown by the Penrose diagrams in square A3 of Fig. (\ref{tablecoll}).


\item {\bf \underline{$ \chi^2+\Lambda > k^2 
\,\,\, , \,\,\, N( \overline r_{in}) < 0 
 \,\,\, ; \,\,\,  P(  \overline r _{out}) < 0  $}}
\\

In the de Sitter diagram the trajectory has the same features
as in the previous case.
In the R-N diagram the shell starts at infinity in a $ I_{-}$ region,
crosses $ r_{+}$ and has two turning points at 
$ r_{out} ^{(1)} $ and $ r_{out} ^{(2)} $ to end in a $ III_{-} $ region.

This is shown by the Penrose diagrams in square B3 of Fig. (\ref{tablecoll}).


\item {\bf \underline{$ \chi^2+\Lambda<k^2 
\,\,\, ; \,\,\, 
 P(  \overline r _{out}) < 0 $}}
\\

In the de Sitter diagram the shell starts at infinity with decreasing radius, 
turns at $ r_{in} $ and reaches the origin in the right patch.
In the R-N diagram the shell behaves as in the previous case.
This is shown by the Penrose diagrams in square B2 of Fig. (\ref{tablecoll}).


\end{enumerate}

We now turn to cases in which there exists some value of $r$ at which
$V(r) > -1$, so that there are inversion points in the trajectory.

\begin{figure}
  \centerline{\epsfbox{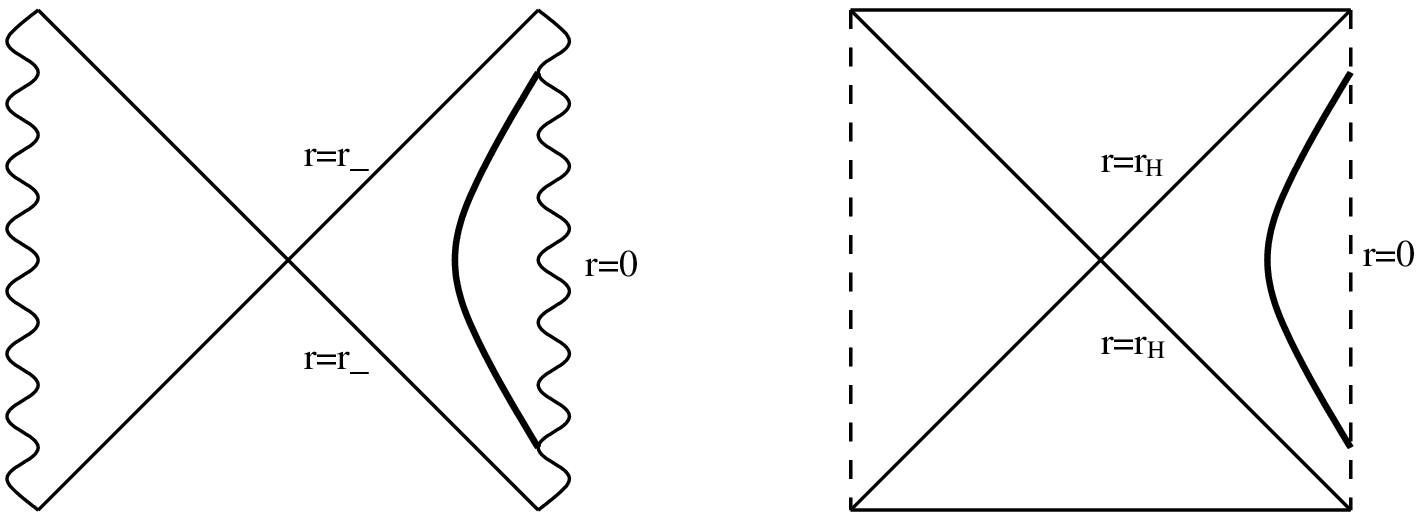}}
  \vspace{0.3in}
  \caption{\sf Penrose diagrams describing the trajectory that begins
    and ends at $r=0$.}
  \label{extrafig}
\end{figure}

For a growing trajectory (starting from zero radius), the shell cannot cross
either the de Sitter horizon $ r_{H} $ or the
inner R-N horizon $ r_{-}$.  Thus,

\bea
   r_{max} & < & r_{-} \nonumber \\
   r_{max} & < & r_{H} ~, 
\eea
and the behavior is represented in Fig. (\ref{extrafig}).

Things are more complicated for collapsing trajectories 
(starting from infinite radius). We will distinguish
different behaviors by the position of the minimum radius 
with respect to the horizons and by the changes in $ s_{in} $ and $ s_{out} $,
so we will describe separately the behavior in the de Sitter
and in the R-N spacetimes, checking then which of these trajectories
can be consistently glued together.

In the de Sitter space there are five possible different behaviors

\begin{figure}
  \hspace{1.5cm}
  \centerline{\epsfbox{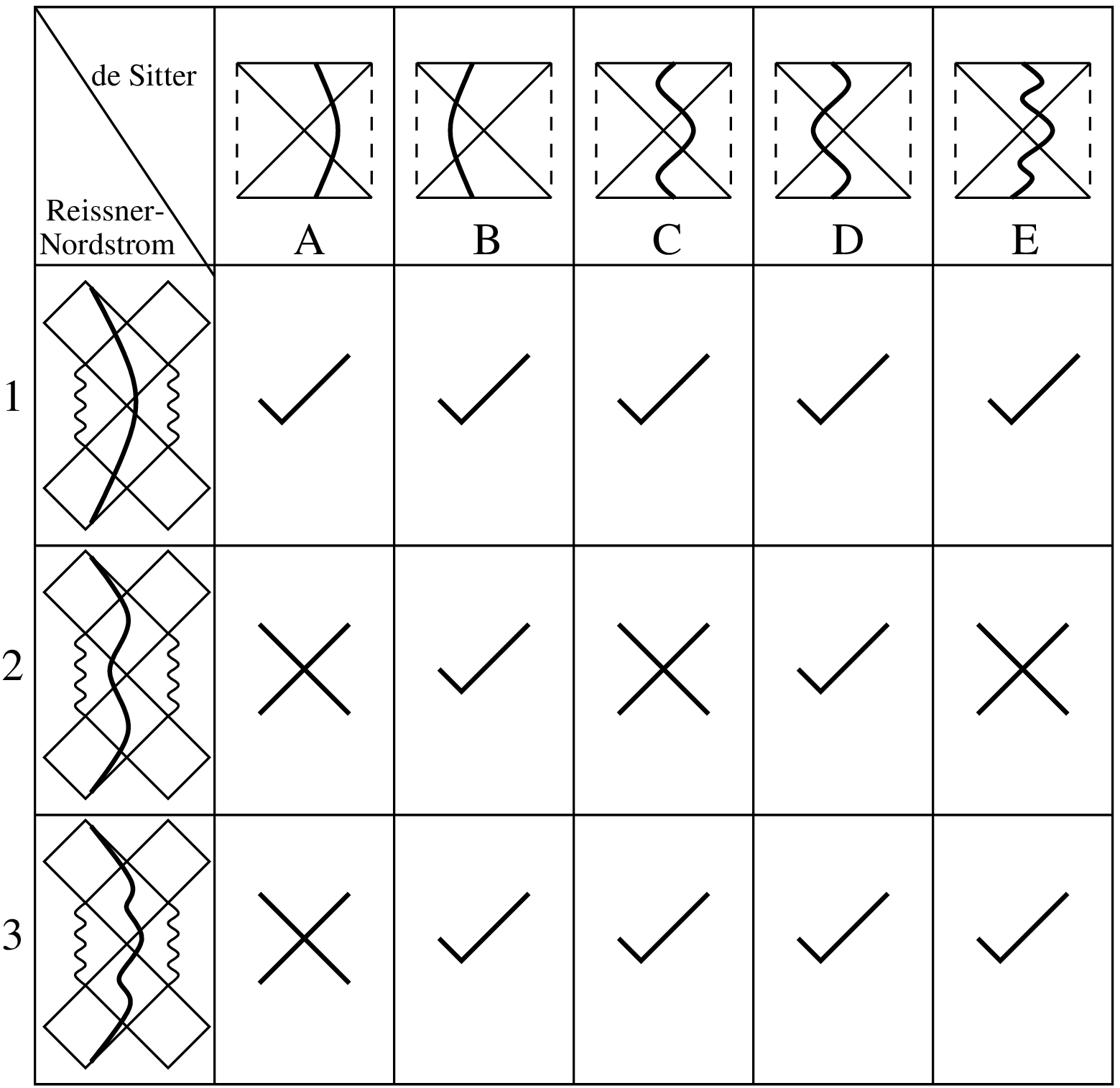}}
  \vspace{0.3in}
  \caption{\sf ``Bounce'' trajectories. Check marks indicate allowed 
trajectories, crosses indicate forbidden trajectories.}
  \label{tablebounce}
\end{figure}

\begin{enumerate}

\item [A)] {\ } 
\be 
s_{in}=-1 
\nonumber 
\ee
for the whole trajectory.

The shell starts from infinite radius,
reaches $ r_{min}  $ and bounces back at infinite radius
always with an increasing angle.
This is shown in square A in Fig. (\ref{tablebounce}).

\item [B)] {\ } 
\be s_{in}=+1  \nonumber 
\ee
for the whole trajectory.
 
The shell starts from infinite radius,
reaches $ r_{min}  $ and returns to infinite radius
always with decreasing angle.
This is shown in square B in Fig. (\ref{tablebounce}).

\item [C)] {\ }
\be  
s_{in}= \left\{\begin{array}{ll}
       +1 & \ \ \ \mbox{$ r_{in} < r $} \\  
       -1 & \ \ \ \mbox{$ r_{min} < r < r_{in}   ~.$}
\end{array}\right. \nonumber
\ee 
The shell starts from infinity with decreasing angle, turns to the right at
$ r_{in} $, reaches $ r_{min} $ with increasing angle and return to infinity
with time reversal behavior. 
This is shown in square C in Fig. (\ref{tablebounce}).

\item [D)] {\ }
\be
s_{in}= \left\{\begin{array}{ll}
 +1 & \ \ \ \mbox{$ r_{min}< r <r_{in} $} \\
 -1 & \ \ \ \mbox{$ r_{in} < r ~.$}
\end{array}\right. \nonumber
\ee 
The shell starts from infinity with increasing angle, reaches 
$ r_{in} $ turning to the left to go to $ r_{min} $ with decreasing
angle and bounces back
at infinity. 
This is shown in square D in Fig. (\ref{tablebounce}).

\item [E)] {\ }
\be
s_{in}= \left\{\begin{array}{ll}
         +1 & \ \ \ \mbox{$ r_{in}^{(1)} < r <r_{in}^{(2)} $}   \\
         -1 & \ \ \ \mbox{$ r_{min} < r < r_{in}^{(1)} \ \ , \ \ 
            r_{in}^{(2)} < r  ~. $}
\end{array}\right. \nonumber
\ee 
The shell starts with increasing angle at infinite radius,
turn to the left to at $ r_{in}^{(2)} $ and then to the right at 
$ r_{in}^{(1)} $ to reach $r_{min}$ and go back to infinity.
This is shown in square E in Fig. (\ref{tablebounce}).

\end{enumerate}

In the R-N space we have three possible behaviors.

\begin{enumerate}

\item[1)]{\ } 
\be s_{out}=-1 \nonumber \ee
for the whole evolution.
 
The trajectory starts from infinite radius in a $ I_{-} $ region and
reaches $ r_{min} $ with decreasing angle in a $ III_{-} $ region to
bounce back at infinite radius in a $ I_{-} $ region.
This is shown in square 1 in Fig. (\ref{tablebounce}).

\item[2)]{\ }
\be
s_{out}=\left\{\begin{array}{ll}
+1 & \ \ \ \mbox{$r_{min} < r<r_{out} $}    \\
-1 & \ \ \ \mbox{$ r_{out}<r ~.$}
\end{array}\right.\nonumber 
\ee 
The shell starts in a $ I_{-} $ region with decreasing angle, turn 
at $ r_{out} $ to reach $r_{min} $ in a $ III_{+} $ region
and bounces back at infinite radius.
This is shown in square 2 in Fig. (\ref{tablebounce}).

\item[3)]{\ } 
\be
s_{out}=\left\{\begin{array}{ll}
+1 & \ \ \ \mbox{$ r_{out}^{(1)}<r<  r_{out}^{(2)}  $}    \\
-1 & \ \ \ \mbox{$ r_{min}<r<r_{out}^{(1)} \ \ , \ \ 
            r_{out}^{(2)} < r ~.  $}
\end{array}\right. \nonumber
\ee 
The shell starts with infinite radius in a $ I_{-} $ region with
decreasing angle, then turns to the left at $ r_{out}^{(1)} $ and 
to the right at $  r_{out}^{(2)} $ to reach $r_{min}$ in a 
$ III_{-} $ region.
This is shown in square 3 in Fig. (\ref{tablebounce}).

\end{enumerate}

For trajectories with $ r_{min} < r_{-}$ it is impossible to glue 
consistent with eq. (\ref{boths}) 
the trajectories A, C and E in the de Sitter diagram
with the 2 in the R-N. Nor can the A trajectory in the de Sitter
diagram be glued with
with the 3 trajectory in R-N.
All the other configurations are possible and have been realized
numerically.

\begin{figure}
  \centerline{\epsfbox{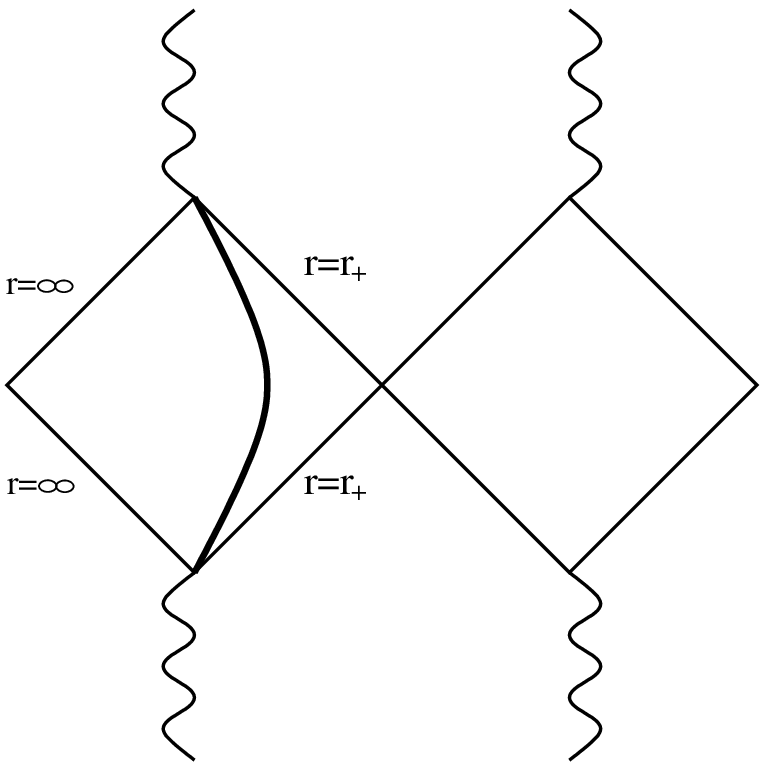}}
  \vspace{0.3in}
  \caption{\sf A bounce trajectory that remains in $I_-$.}
  \label{figlast}
\end{figure}

The only trajectories remaining are those with $ r_{min}> r_{+} $.
In the R-N conformal diagram they will always stay in the $ I_{-} $ region
as we show in Fig. (\ref{figlast}).
This can be glued consistently with the A, B, and D trajectories in the 
de Sitter spacetime.

The bubble solutions we have described in this section contain timelike
singularities. It is interesting
to consider whether these solutions could be patched onto
solutions with smooth initial data, removing the past timelike
singularities.
Physically, this would correspond
to the ability to build a machine that would be capable of producing a 
black hole with an inflating interior.

Uncharged collapsing shells have previously been considered
in \cite{BGG}. The problem encountered there was that whenever the bubble grew
sufficiently large, anti-trapped regions  would necessarily appear
\cite{FG 87}. By the Penrose singularity theorem this implies a past
directed null geodesic must be incomplete \cite{FG 87}. This amounts to the 
condition that initial conditions must be singular.

Similar solutions have been considered in the context of colliding
monopoles in \cite{BTV}. There a way of circumventing the problem of
singular initial data was suggested. A timelike singularity would form 
behind an event horizon, and the anti-trapped regions that must
necessarily form, would be to the future of this timelike singularity.
In this way, the Penrose theorem is satisfied, and it can be
possible to realize this solution with smooth initial data.

This is exactly the state of affairs in the collapse and bounce bubble
solutions we have constructed. In principle, therefore, it should be
possible to patch these bubble solutions onto smooth initial
data. However, another obstacle encountered in \cite{BTV}
was that classically it is
not possible to evolve the solutions past the timelike singularity
without specifying additional boundary conditions on the
singularity. 
In the next section, we describe how this problem is solved for analogous
asymptotically anti-de Sitter bubble solutions by appealing to the
duality between gravity in AdS backgrounds and large $N$ conformal
field theory.

\section{AdS/CFT Interpretation}
\label{sec:adscft}

In principle, the correspondence between gravity in an anti-de Sitter background 
and conformal field theory on the boundary \cite{maldacena}, allows us to 
map any process in the gravity theory to a statement in the
field theory. 
The duality is conjectured to hold for any spacetime dimension bigger
than 2 (on the gravity side).
The simplest case to consider is five-dimensional
supergravity in an AdS background which is dual to four-dimensional 
$SU(N)$ Yang-Mills
theory with 4 supersymmetries. The radius of curvature $R$ of AdS
and the string coupling $g_s$ are
related to parameters in the Yang-Mills theory by
\be
R=(g_{YM}^2 N)^{1/4}~, \qquad g_s = g_{YM}^2~.
\ee
The
large $N$ limit, with $R$ fixed, will correspond to the classical
limit of the supergravity theory.  For other dimensions,
the CFT is an exotic superconformal field theory which is difficult to 
describe explicitly.
The important point for us is that
evolution in the CFT is unitary, which
allows us to describe, in a completely well-defined way, 
processes in gravity which lead to classical
singularities. 

The bubble solutions we have discussed, and ones previously found
\cite{BGG},
carry over to backgrounds where the asymptotically flat backgrounds 
are replaced by asymptotically anti-de Sitter backgrounds. Furthermore 
they can be generalized to general spacetime dimensions (the relevant
black hole solutions may be found in \cite{myers}.) The
qualitative properties of the trajectories will not change provided
$R$ is not taken to be too small relative to the other parameters. 
We will take
the point of view that the correspondence between gravity 
and conformal field theory is valid, and discuss what these solutions
must correspond to from the CFT point of view. 

The AdS/CFT conjecture is most precisely formulated in Euclidean
space \cite{witten}. We will follow the procedure discussed in
\cite{lowelarus}, 
to obtain 
a definition of the Lorentzian gravity theory. The logic is as
follows:
the conjecture asserts that all observables in the Euclidean gravity
theory can be computed in terms of gauge invariant observables in the
Euclidean gauge theory. In particular, we assume this gives us enough
information to construct the metric in the gravity theory (modulo
coordinate transformations), for any given process in the gauge
theory. The Lorentzian gauge theory is defined by the usual Wick
rotation from imaginary time in the correlation functions, which leads 
to a definition for the observables in the Lorentzian gravity theory
in terms of a Wick rotation of the Euclidean observables.

Reissner-Nordstrom black holes have been previously considered in the
AdS/CFT context by Chamblin, Emparan, Johnson and Myers \cite{myers}. 
In terms of
the CFT, the black hole microstates correspond to a finite temperature
plasma of charge associated with a $U(1)$ subgroup of the
global R-symmetry. 

Let us consider the interpretation of the bubble trajectories in terms 
of the large $N$ CFT. A bubble in region I far from the event horizon
will correspond to a charged shell in CFT. The description of analogous
neutral shells in the AdS/CFT correspondence has recently been studied 
by \cite{itzhaki,PST}. For spherically symmetric shells, the radial
position is encoded in non-local correlations in
the CFT state.
For the cases of most interest for us,
the allowed collapse and bounce trajectories of the
charged bubbles are always behind the event horizon of a black hole
from the point of view of an outside observer. These solutions
therefore correspond to particular black hole microstates from the
point of view of an outside observer. Thus in the CFT description, the
bubbles correspond to
particular microstates in the finite temperature plasma 
associated with AdS Reissner-Nordstrom black holes. It is difficult to 
describe these states explicitly without more direct control over 
calculations in the strongly coupled gauge theory.
If it is possible to indeed patch the bubble solutions onto
configurations with smooth initial data, it would be much more
straightforward to map this into initial data in the CFT, using the
current understanding of the mapping between states in the CFT to
states in the gravity theory \cite{witten}.

The interior of the domain wall is de Sitter space. In order for this
to be a solution of the gravity theory, we suppose that appropriate
perturbations are turned on in the CFT to lead to a phase with broken
supersymmetry inside the domain wall. A non-trivial potential may then 
be generated for one of the scalars in the gravity theory, allowing it 
to act as an inflaton field, generating a non-trivial vacuum energy in 
this region.

It is possible to construct a unitary
time evolution operator which evolves states across the event horizon  
and through the singularity by combining the usual field theory time
translation operator with the conformal generators. In the case of a
neutral black hole in
$AdS_3$ this construction has been made explicit \cite{banks}.
Thus the CFT description resolves the classical
singularity of the black hole geometry and leads to a prescription
for evolving gravity solutions to the future of timelike
singularities. The CFT implies definite
boundary conditions on the timelike singularities
which in general will depend on the initial state. In the large $N$
limit, and for sufficiently large bubbles, 
it should be possible to make these conditions arbitrary by an
appropriate choice of the CFT initial state. This follows from the
fact that an arbitrary configuration in the gravity theory can
be mapped to a configuration in the CFT. 
The bubble solutions described
here correspond to boundary conditions fixed by analyticity of the
metric.

Finite $N$ corrections in the CFT correspond to quantum corrections
from the gravity point of view. Black holes will be
unstable at the quantum level
due to Hawking radiation. The supergravity theory contains
charged, light fields, hence the Hawking radiation will tend to discharge
the black hole \cite{damour}.
In line with the discussion of black hole
complementarity in AdS/CFT of \cite{lowelarus}, the CFT implies the
following picture for the quantum evolution of the bubble in the
gravity theory. The gravity variables inside and outside the event
horizon are redundant. Outside the horizon, the final state will be a
set of light outgoing quanta which carry all the information about
processes inside the black hole event horizon. This will now be a
complicated time dependent state from the CFT viewpoint. The degrees of freedom
inside the horizon, including those describing the eternally inflating 
region, are encoded in the same field theory degrees of freedom that
describe physics outside.

Note the most interesting bubble
solutions contain eternally 
inflating regions. 
The description 
of these solutions in terms of the CFT implies that even though we
have a large (and increasing) proper volume inside the bubble, only a
finite number of states can exist in this region. The
Bekenstein-Hawking entropy
of the black hole is finite, and will bound the number of 
field theory microstates corresponding to such configurations.

The picture of the quantum evolution of the bubble described above alleviates
another problem with the classical solution we have thus far not
addressed. A Cauchy horizon is classically unstable  to small
perturbations, which grow producing weak curvature singularities in the
vicinity of the Cauchy horizon \cite{poisson,ori}. A free-falling observer
crossing the Cauchy horizon sees the entire history of the external
universe in the moments prior to crossing. The picture of the
quantum evolution of our solutions we have discussed above solves
this problem. 
The black hole evaporates in a finite time, which
avoids the problem of an observer inside seeing the entire history
of the external universe. Furthermore, because the CFT description
gives a resolution of the classical singularity of the black hole, the 
Cauchy horizon is no longer present quantum mechanically.

What is less clear is precisely how 
this singularity is resolved. One possibility is that a spacelike
region of strong curvature forms outside the would-be Cauchy horizon,
the system behaves in the same way as a Schwarzschild black hole,
and the ``tunnel'' to the inflating region collapses. 
Another possibility is that the curvature remains finite in the
vicinity of the inner horizon, and we do indeed have an eternally
inflating region inside the bubble. We propose,
at least for special choices of
initial conditions, that the second option is viable. It seems likely
that the constraints on these initial conditions are very strong, and
it would be
interesting to study further the experimental feasibility of
creating an inflating universe in the laboratory.

\section*{Acknowledgments}

We would like to thank Stephon Alexander, Roberto Balbinot, 
Cyrus Taylor and Tanmay Vachaspati for helpful 
conversations.  The work of M.T. was supported by the Department of Energy (D.O.E.).
The research of D.L. is supported in part by DOE grant DE-FE0291ER40688-Task A.

\end{document}